# Generic configuration stellarator based on several concentric Fourier windings

**V. Queral**

*Laboratorio Nacional de Fusión, CIEMAT, 28040 Madrid, Spain*
E-mail: vicentemanuel.queral@ciemat.es

**Abstract**
Stellarators commonly comprise different sets of coils to produce diverse magnetic configurations. However, the diversity of possible configurations in a single device is usually rather limited. The achievement of a broad variety of magnetic configurations might be valuable for some purposes, for example, to assay the effect of the magnetic configuration on turbulent transport. Thus, a method is created to systematically define sets of modular coils located on concentric toroidal winding surfaces. The method is based on the expression of a Last Closed Flux Surface (LCFS) by Fourier coefficients in cylindrical coordinates and consists in the definition of successive windings located on equidistant concentric winding surfaces, each winding such that produces a magnetic field which, when added to the magnetic field generated by a sole base winding that generates a base magnetic configuration, produces a magnetic configuration whose LCFS is defined by the Fourier coefficients of the base magnetic configuration plus only one new Fourier coefficient. The utilization of a diversity of currents in the different windings would give a linear combination of magnetic fields that reproduce the LCFS defined mathematically by the respective Fourier coefficients. The deviation between a particular modelled LCFS and the obtained LCFS from the windings depends on: the order and value of each Fourier coefficient, the selected shape of concentric winding surfaces, and the possible intersection of the LCFS with the internal winding surface.
The method to generate the windings is reported and the application to one case study is described. Each modelled LCFS is compared with the respective Poincaré plot obtained from the linear combination of magnetic fields from the windings.

**1. Introduction**
Commonly stellarators comprise different sets of coils to produce diverse magnetic configurations. For example, HSX stellarator [1] is equipped with a set of modular coils and a set of planar coils to modify the magnetic configuration. However, the diversity of magnetic configurations in a single device is limited by the number of windings and the method to define the windings. The term *winding* is referred next to a set of twisted modular coils located on a toroidal winding surface, each winding as described in Ref. [2]. An arbitrary number of windings on concentric winding surfaces ([3]-p.34) can be calculated by the developed method.

Section 2 defines the two phases stablished to generate the windings. The application of the method to a case study is described in Section 3. Section 4 compares the intended LCFS with the Poincaré plots from the magnetic field obtained from the series of windings and evaluates the applicability limits of the method. Section 5 outlines a similar inverted method.

**2. Definition of the method**
The stellarator is assumed having stellarator symmetry and periodic. Also, the magnetic configurations are considered in vacuum, without the effect of the plasma pressure or bootstrap current. Therefore, a LCFS defines a single magnetic configuration [2]. A LCFS of such stellarator can be expressed by a Fourier series in cylindrical coordinates [2].

$$r = \sum_{m=0, n=-n_b}^{m_b, n_b} R_{mn} \cos 2\pi (mu + nv) \qquad z = \sum_{m=0, n=-n_b}^{m_b, n_b} Z_{mn} \sin 2\pi (mu + nv) \qquad (1)$$

being:

($r, \varphi, z$) cylindrical coordinates.   $\varphi = (2\pi v)/n_p$
$u, v \in [0,1]$ angle like variables.
$n_p$ number of periods of the stellarator.
$R_{mn}$ and $Z_{mn}$ Fourier coefficients. $m$ and $n$ are the poloidal and the toroidal mode number respectively.

*2.1 Notation and description of the method*

$\mathbf{C}_b$ and $\mathbf{C}_k$ denote a magnetic configuration. LCFS$_b$ and LCFS$_k$ denote a Last Closed Flux Surface. $\mathbf{S}_b$ and $\mathbf{S}_k$ denote a winding surface and $\mathbf{W}_b$ $\mathbf{W}_k$ a winding. The subindex 'b' denotes a *base* element and 'k' the *k-th* element.

Two phases are followed in order to define the series of windings.

**1$^{st}$ phase.** Calculation of a first winding, named *base winding*, which produce a *base (magnetic) configuration*. In general, the base configuration will present smooth contours defined by low order Fourier coefficients in expression (1).

The base winding is generated similarly to a common process to generate modular coils ([4]-p.859) for stellarators and explained elsewhere [2, 5]. However, some particularities have to be considered for the present method since a broad variety of magnetic configurations are aimed. In particular, the Fourier coefficients $R^b_{mn}$ and $Z^b_{mn}$ defining the base configuration should be preferably of low order since the second phase of the method seeks to create windings based on Fourier coefficients of higher order, for the detailed shaping of the configuration. Thus, the base configuration $\mathbf{C}_b$, generated by a *base winding* $\mathbf{W}_b$, is defined by the expression (1) by non-zero low order Fourier coefficients { $R^b_{mn}$ , $Z^b_{mn}$ }

**2$^{nd}$ phase.** Calculation, by the procedure described next, of several concentric additional windings $\mathbf{W}_k$, named *Fourier windings*, located on equidistant concentric winding surfaces. The term *additional winding* means a non-base winding.

Each additional winding $\mathbf{W}_k$ is such that generates a magnetic field, which added to the magnetic field created by the base winding $\mathbf{W}_b$, generates a magnetic configuration $\mathbf{C}_k$ whose LCFS is defined by the Fourier coefficients of the base configuration <u>plus only one new Fourier coefficient</u>. That is,

$$\mathbf{B}(\mathbf{C}_k) = \mathbf{B}(\mathbf{W}_b) + \mathbf{B}(\mathbf{W}_k) \qquad k = 1\ldots N\text{-}1 \qquad (2)$$

$\mathbf{C}_k$ defined by { $R^b_{mn}$ , $Z^b_{mn}$ } ∪ (one $R^k_{mn}$ or one $Z^k_{mn}$ )

being:

$N$ : Number of windings including the base winding.

$\mathbf{B}$ : Magnetic field generated by a particular winding or, which generates certain magnetic configuration.

$R^k_{mn}$ , $Z^k_{mn}$ : Arbitrary Fourier coefficients not included in $\mathbf{C}_b$.

Each winding surface $\mathbf{S}_b$ $\mathbf{S}_k$ is defined equidistant to one single *reference winding surface,* $\mathbf{S}_{ref}$. Nevertheless, equidistance is not a necessary condition for the generation of the coils. Also, the condition of only one additional new Fourier coefficient is not strictly necessary but allows the generality of the method.

It is assumed linearity between the currents in the windings and the effect on the particular Fourier coefficient which define the LCFS. The next section try to elucidate, from one particular case study, up to what level the magnetic field from the windings replicate the intended LCFS obtained from (1).

## 3. Application of the method to a case study

*3.1 Limits for the calculation of windings*

Any generic magnetic configuration $\mathbf{C}_k$ in expression (2) cannot be exactly generated by distant coils from the LCFS. This limitation is discussed in Ref. [6]-p.1102. For the generalised configuration stellarator, $\mathbf{S}_k$ is not equidistant to LCFS$_k$, in general. Therefore, the distance from $\mathbf{S}_k$ to a particular LCFS$_k$ and other LCFS is variable. Moreover, the distance cannot be negligible since all the possible LCFS from (1) should be interior to the internal winding surface, Figure 1. Thus, some of the LCFS in Figure 1 would not be accurately reproduced by the concentric windings.

*3.2 Selection of a reference winding surface*

A reference winding surface $\mathbf{S}_{ref}$ is shaped as an average of most of the possible LCFS of interest in the device. $\mathbf{W}_{N-1}$ is taken as the internal winding surface since it corresponds to the highest order Fourier coefficients, whose effect decrease faster with distance [7].

*3.3 Definition of the Fourier coefficients*

NESCOIL code [2] calculates a set of modular coils from a winding surface and a background field generated by another set of coils ($W_b$ in this case). The modelled stellarator is of 3 periods and $N = 7$.

The low order Fourier coefficients in Table 1 are taken for the base configuration $C_b$. For simplification, the magnetic axis is chosen planar and approximately circular but extra $R_{0n}$ $Z_{0n}$ coefficients may be included in the base configuration to shape the magnetic axis of $C_b$. Table 1 shows also the coefficients for $S_{ref}$. Figure 2 illustrates a poloidal cut of the $LCFS_b$ and $S_b$.

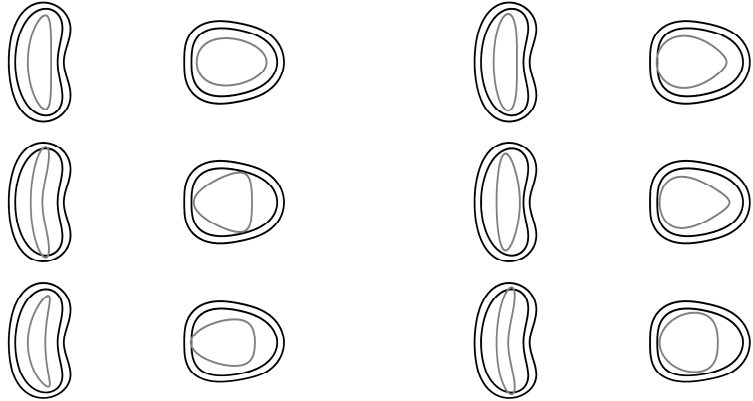

**Figure 1.** Poloidal cuts of some common and some abnormal LCFS (grey) and external and internal winding surfaces (black) for the case study.

| $m$ | $n$ | $R^b_{mn}$ | $Z^b_{mn}$ | $S_{ref}\ R_{mn}$ | $S_{ref}\ Z_{mn}$ |
|---|---|---|---|---|---|
| 0 | 0 | 4.0 | 0.0 | 4.0 | 0.0 |
| 1 | 0 | 1.0 | 1.5 | 1.0 | 1.5 |
| 1 | 1 | 0.5 | -0.5 | 0.45 | 0.45 |
| 2 | 0 | | | 0.25 | 0.0 |

**Table 1.** Fourier coefficients for the base configuration $C_b$ and for $S_{ref}$.

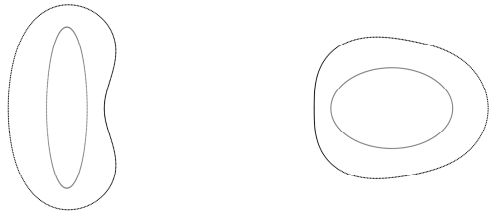

**Figure 2.** Poloidal cut of the $LCFS_b$ (grey) and $S_b$ (black).

The Fourier coefficients ($R^k_{mn}$ and $Z^k_{mn}$) selected for the successive $C_k$ configurations are shown in Table 2. The selection of the particular Fourier coefficients in Table 2, the assigned values and sign are stablished from the knowledge of other magnetic configurations (e.g. a quasi-isodynamic of 3 periods [8]), the author's previous experience [3] and tries. The mean error from NESCOIL should be lower than 1% [9] and the maximum error ~4%. For example, a Fourier coefficient $R^3_{21} = -0.25$ (Table 2) gave a mean error of 2% on $S_3$. Nevertheless, appropriate coils resulted with $R^3_{21} = -0.12$. Higher order Fourier coefficients allow in general lower value according to the limits cited in Section 3.1.

| $k$ | $m$ | $n$ | $R^k_{mn}$ | $Z^k_{mn}$ | $R_{mn} = \lambda_k R^k_{mn}$ | $Z_{mn} = \lambda_k Z^k_{mn}$ |
|---|---|---|---|---|---|---|
| 1 | 2 | 0 | 0.25 | | $\lambda_1\ 0.25$ | |
| 2 | 2 | 0 | | 0.25 | | $\lambda_2\ 0.25$ |
| 3 | 2 | 1 | -0.12 | | $\lambda_3\ (-0.12)$ | |
| 4 | 2 | 1 | | -0.25 | | $\lambda_4\ (-0.25)$ |
| 5 | 1 | 2 | -0.06 | | $\lambda_5\ (-0.06)$ | |
| 6 | 1 | 2 | | 0.12 | | $\lambda_6\ 0.12$ |

**Table 2.** Fourier coefficients for the $C_k$ configurations in arbitrary units.

*3.4 Calculation procedure for the windings*

7 poloidal and 8 toroidal modes are used in NESCOIL code. Fourier coefficients of the potential [2] are obtained from NESCOIL for each winding.

• $W_b$ is calculated from $C_b$ and $S_b$, being $S_b$ equidistant 0.7 units from the reference winding surface $S_{ref}$.

• Calculation of the additional winding $W_1$

$C_1 \equiv C_b \cup R^1_{20}$ ($R^1_{20}$ Table 2), $S_1$ is externally equidistant 0.65 units from the reference winding surface $S_{ref}$, and the background magnetic field from the base winding $W_b$ set to two units of current.

The winding surface $S_b$ for $W_b$ is the exterior one. The next winding surfaces $W_k$ are internally equidistant to $S_b$.

• $W_k$ is calculated as $W_1$ following the general expression (2). The distance from $S_k$ to $S_{ref}$ is $d = 0.7 - 0.05\,k$ units, being $S_k$ external to $S_{ref}$.

*3.5 Calculation of Poincare plots from the windings*

CASTELL code [3] is utilised for the next calculations. Windings are obtained from the Fourier coefficients of the potential [2] from NESCOIL. The magnetic field at each point of N three-dimensional grids, one for each winding, is calculated by Biot–Savart law. Additionally, CASTELL code generates an N-dimensional discrete space of $P$ different $\lambda^i_k$ values. The linear combination of magnetic grids is calculated for each of the $P$ points in the discrete space and stored.

$$\mathbf{B}_i = \mathbf{B}(\mathbf{W}_b) + \sum \lambda^i_k \mathbf{B}(\mathbf{W}_k) \qquad k = 1 \text{ to } N\text{-}1 \qquad i = 0\ldots P\text{-}1$$
$$0 \leq \lambda_k \leq 1 \qquad \lambda^i \equiv (\lambda^i_1, \ldots, \lambda^i_{N-1})$$

Poincaré plots are generated by guiding centre orbit integration by CASTELL.

The number of elements of the discrete space is hardly manageable if $N$-1 = 6 additional windings and several points in the interval $0 \leq \lambda_k \leq 1$ are considered. Thus, only 4 windings plus the base winding are included next. Also, only 3 points $\lambda_k \in \{0.0, 0.5, 1.0\}$ are taken. It results $P = 81$ different magnetic configurations.

Figure 3 shows the obtained $i$ Poincaré plots. The target LCFS calculated by expression (1) is represented in grey. The last configuration $i = 80$ is not shown.

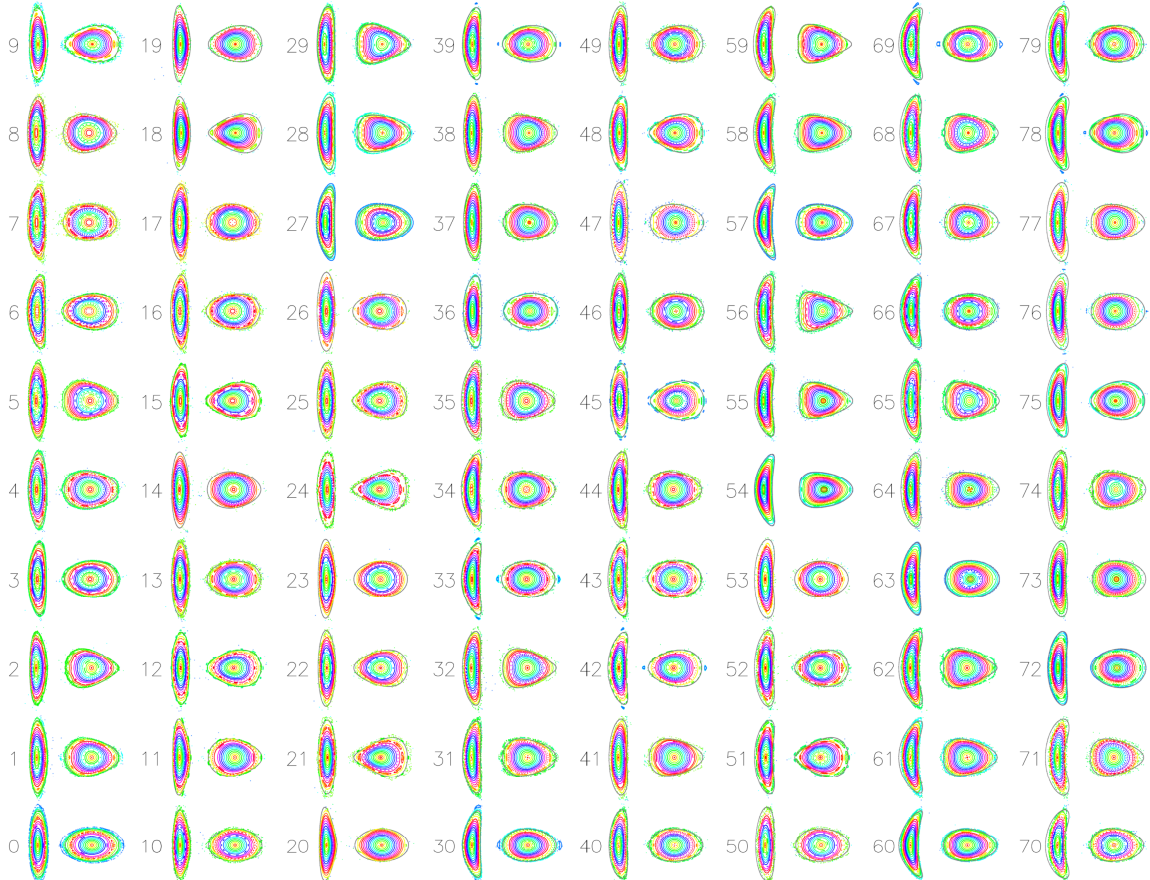

**Figure 3.** Poincaré plots for 80 configurations from combination of 5 windings and 3 currents per winding. The intended LCFS is represented in grey.

$\lambda_k = 1$ corresponds to a unit of current in the winding.

Table 3 lists 20 *n*-tuples $\lambda^i$ of the total 81 *n*-tuples and the mean error (ME) $\mathbf{B}\cdot\mathbf{n}$ on each $\text{LCFS}^i$ from CASTELL.

| i | $\lambda^i_1$ | $\lambda^i_2$ | $\lambda^i_3$ | $\lambda^i_4$ | ME | i | $\lambda^i_1$ | $\lambda^i_2$ | $\lambda^i_3$ | $\lambda^i_4$ | ME |
|---|---|---|---|---|---|---|---|---|---|---|---|
| 0 | 0 | 0 | 0 | 0 | 0.1 | 70 | 1 | 0.5 | 1 | 0.5 | 3.4 |
| 1 | 0 | 0 | 0 | 0.5 | 0.6 | 71 | 1 | 0.5 | 1 | 1 | 4.8 |
| 2 | 0 | 0 | 0 | 1 | 0.2 | 72 | 1 | 1 | 0 | 0 | 1.6 |
| 3 | 0 | 0 | 0.5 | 0 | 0.3 | 73 | 1 | 1 | 0 | 0.5 | 2.3 |
| 4 | 0 | 0 | 0.5 | 0.5 | 0.2 | 74 | 1 | 1 | 0 | 1 | 3.4 |
| 5 | 0 | 0 | 0.5 | 1 | 1.1 | 75 | 1 | 1 | 0.5 | 0 | 2.4 |
| 6 | 0 | 0 | 1 | 0 | 0.1 | 76 | 1 | 1 | 0.5 | 0.5 | 3.3 |
| 7 | 0 | 0 | 1 | 0.5 | 0.8 | 77 | 1 | 1 | 0.5 | 1 | 4.4 |
| 8 | 0 | 0 | 1 | 1 | 2.0 | 78 | 1 | 1 | 1 | 0 | 3.3 |
| 9 | 0 | 0.5 | 0 | 0 | 0.4 | 79 | 1 | 1 | 1 | 0.5 | 4.3 |

**Table 3.** $\lambda^i_k$ values for the calculation of some of the configurations appearing in Figure 3.

*3.6 Exploration of higher order coefficients*

A higher order Fourier coefficient $R^7_{40} = 0.06$ and a new winding is added to the previous data in Table 2 and Table 3. Table 4 shows the inputs and resulting mean error. Figure 4 displays the Poincaré plots.

| i' | $\lambda^i_1$ | $\lambda^i_2$ | $\lambda^i_3$ | $\lambda^i_4$ | $\lambda^i_5$ | $\lambda^i_6$ | $\lambda^i_7$ | ME |
|---|---|---|---|---|---|---|---|---|
| 0' ≡ i=3 | 0 | 0 | 0.5 | 0 | 0 | 0 | 0 | |
| 1' | 0 | 0 | 0.5 | 0 | 0 | 0 | 0.5 | 0.2 |
| 2' | 0 | 0 | 0.5 | 0 | 0 | 0 | 1 | 0.4 |
| 3' ≡ i=4 | 0 | 0 | 0.5 | 0.5 | 0 | 0 | 0 | |
| 4' | 0 | 0 | 0.5 | 0.5 | 0 | 0 | 0.5 | 0.7 |
| 5' | 0 | 0 | 0.5 | 0.5 | 0 | 0 | 1 | 1.2 |
| 6' ≡ i=5 | 0 | 0 | 0.5 | 1 | 0 | 0 | 0 | |
| 7' | 0 | 0 | 0.5 | 1 | 0 | 0 | 0.5 | 1.7 |
| 8' | 0 | 0 | 0.5 | 1 | 0 | 0 | 1 | 2.4 |

**Table 4.** Inputs and result for the calculation of the configurations appearing in Figure 4.

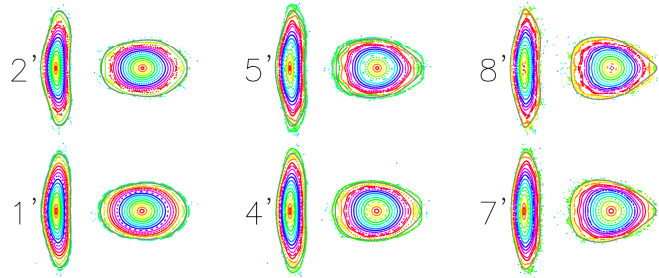

**Figure 4.** Poincaré plots for six configurations including one high order Fourier coefficient.

**4. Comparison of the LCFS and the Poincare plots from the windings**

The next conclusions and results are obtained from the calculations and Figure 3.

A. Most of the magnetic surfaces from the Poincaré plots reproduce acceptably the respective LCFS. In general, the mean error (Table 3) is lower than 3%, considered appropriate for the objective of generating a broad diversity of magnetic configurations.

B. Configuration 2, 6, 18 and 54 are generated by the base winding and only one additional winding. Small magnetic islands appear near the edge of the configurations. They obey to the rational 3/8 for the configurations 2, 18 and 54.

C. Rational surfaces in the plasma cannot be avoided for many configurations since they are generated automatically. For example, configurations 2 and 11 contain the rational 1/3 and many configurations contain the rational 3/8. The rotational transform ranges from 0.25 at the plasma centre for some configurations to 0.5 at the edge for others. A careful design of the base and additional configurations would reduce the number of low order rationals in the plasma. However, rationals might even have positive effect under certain circumstances [10].

D. The agreement of the 'egg shape' of the LCFS with the poloidal cut of the Poincaré plot is satisfactory in most of the configurations except for sharp LCFS like e.g. configurations 56, 59, 62, 65. The mean error in such configurations is high. The agreement of the 'bean shape' of the LCFS is unsatisfactory for some configurations. It is discussed in the next point.
**E.** The indentation is not correctly reproduced in the Poincaré plot in e.g. configurations 61, 62, 70, 71, 75-79. Two hypothesis for such behaviour are: **i)** The indentation of the winding surfaces (Figure 2) is modest. Concavities in a LCFS are difficult to generate by distant coils if the concavity is not replicated at the winding surface. Nevertheless, configuration 31, 33, 54, 57, 60 and others reproduce the indentation acceptably well, **ii**) The assumed linearity between the currents in the windings and the effect on the LCFS may not be well fulfilled for combined low order (high value) and high order (low value) Fourier coefficients, or when many Fourier coefficients of large value are combined.
**F.** The inclusion of higher order coefficients should produce smaller but finer details on the LCFS shape. One example has been calculated in Section 3.6. The mean error (Table 4) is reasonably low for most of the configurations.

## 5. Inverted method

An inverted method can be conceived. It consists on generating a configuration $C_k$ by one winding $W_k$

$$B(C_k) = B(W_k)$$

and obtaining the total field by subtracting the base magnetic field from the field generated by each winding $W_k$. That is,

$$B_i = (1 - \sum \lambda^i_k) B(W_b) + \sum \lambda^i_k B(W_k)$$

The same process in Section 3.5 was followed. The mean errors and deviations between the LCFS from (1) and the Poincaré plots from the windings are similar to the mean errors and deviations described in Section 4.

The difference with respect the previous method resides essentially in the current in the base winding, which might be high in the inverted method for the addition of numerous Fourier coefficients of high value.

## 6. Results, conclusions and future work

A method to calculate concentric windings for modular stellarators in a systematic manner has been developed and satisfactorily assayed in a particular case.

The method is based on the incorporation of a single Fourier coefficient of a LCFS and the calculation of a winding by subtracting the magnetic field of a base configuration from the intended magnetic configuration.

A series of 81 configurations have been generated from 5 windings. Most of the magnetic surfaces from the Poincaré plots reproduce correctly the respective LCFS.

A broad range of configurations have been obtained in a single modelled device.

Considerable work remains to be done, for example, the study of the achievable indentation given a particular reference winding surface, the effect of cumulative Fourier coefficients of high value, and the capability of the method to cope with a generic magnetic axis. The feasibility of building a real experimental stellarator equipped with 5 or 10 concentric windings is also debatable and should be advanced. Perhaps, one of the few reasonable methods to create such real experimental device may be additive manufacturing [11]. Also, the location of ports for diagnostics, vacuum pumping and heating would be arduous.

Such an experiment might be valuable to advance still open questions, like for example the effect of the magnetic configuration on turbulent transport.

**Acknowledgement**
This work was partly funded by the Spanish 'Ministry of Economy, Industry and Competitiveness' under the grant number ENE2015-64981-R (MINECO/FEDER, EU). The author would like to thank Dr. Harry Mynick for suggesting time ago a stellarator capable to generate configurations between NCSX and a potentially turbulence improved NCSX configuration, and Dr. Edilberto Sánchez since, after a conversation about the huge processor time needed to calculate turbulent transport, I envisaged the sort of analogic 'computer' developed in this work.